# Estimation of aboveground biomass in a tropical dry forest: An intercomparison of airborne, unmanned, and space laser scanning

Nelson Mattié[1*], Arturo Sanchez-Azofeifa[1], Pablo Crespo-Peremarch[2], Juan-Ygnacio López-Hernández[3]

[1] Centre for Earth Observation Sciences, Department of Earth and Atmospheric Sciences, University of Alberta, Edmonton, AB T6G 2E3, Canada. *Corresponding author: Nelson Mattié, mattiedj@ualberta.ca

[2] Geo-Environmental Cartography and Remote Sensing Group (CGAT), Department of Cartographic Engineering, Geodesy and Photogrammetry, Universitat Politècnica de València, Camí de Vera s/n, 46022 València, Spain.

[3] Photogrammetry and Remote Sensing Laboratory, Faculty of Forestry and Environmental Sciences, Universidad de Los Andes, Mérida, Venezuela.

## Abstract

Accurate aboveground biomass (AGB) estimates are needed to report forest carbon, yet tropical dry forests (TDFs) remain poorly studied. We compared airborne (ALS), unmanned (ULS), and simulated spaceborne full-waveform (SLS, GEDI simulator) laser scanning for AGB estimation in ten permanent plots of the Santa Rosa National Park Environmental Monitoring Super Site, Costa Rica, spanning early to late succession. Reference AGB (22–220 Mg ha-1) was computed from the 2021 census, screening diameter time series, propagating measurement, wood density, and allometric errors (plot uncertainty 3.5–14.5%). We compared two ordinary least squares (OLS) models, a log–log model of mean canopy height fixed a priori and a single metric selected from the data, with support vector machine regression (SVM). Predictor selection, scaling, and tuning were nested within leave-one-plot-out validation. ALS gave the lowest errors: the a priori OLS model reached an RMSE of 32.3 Mg ha-1 (29%; $R2_{LOO}$ = 0.77) and SVM 35.8 Mg ha-1 (32%). SVM did not improve on OLS for any platform and was significantly worse with ULS ($p$ = 0.01). The best ULS and SLS models reached 33% and 37%, respectively; platform differences were not significant with ten plots. Selecting predictors on all plots before cross-validation lowered the apparent SVM error by up to 23 percentage points, and non-structural simulator outputs produced spurious skill, explaining the higher accuracies reported in an earlier version. Model errors were at least 2.5 times the field-AGB uncertainty. With small plot networks, a height-based OLS model with ALS is the most robust choice.

## 1. Introduction

Aboveground biomass (AGB) is a key variable in the terrestrial carbon cycle, and its estimation supports ecosystem modelling, conservation policy, and climate-change mitigation (Houghton et al., 2009). Synthetic aperture radar, passive optical sensors, and laser scanning have all been used to estimate AGB, but laser scanning yields the most accurate models because it directly measures the three-dimensional structure of the canopy (Zolkos et al., 2013). Laser scanning data differ by platform, airborne (ALS), unmanned (ULS), terrestrial, mobile, and spaceborne (SLS), and by recording format, discrete-return or full-waveform (Crespo-Peremarch et al., 2020). Platform characteristics such as trajectory, scanning geometry, footprint size, and point density determine the forest metrics that can be derived, as shown for carriage-based scanning in operational forestry (Gollob et al., 2023) and for small- versus large-footprint lidar in tropical forests (Silva et al., 2018). The Global Ecosystem Dynamics Investigation (GEDI), which has been operating from the International Space Station since 2019, now provides full-waveform footprints of approximately 25 m designed for vegetation structure and biomass (Dubayah et al., 2020; Duncanson et al., 2022).

Tropical dry forests (TDFs) are the most threatened tropical biome and store a significant but poorly quantified fraction of tropical carbon (Sánchez-Azofeifa et al., 2005; Portillo-Quintero et al., 2015). Their deciduous, low, and heterogeneous canopies and mosaics of successional stages differ from those of humid forests, in which most laser-scanning biomass models were developed (Kalacska et al., 2004; Sun et al., 2019; Zhao et al., 2021).

### 1.1. The multi-platform intercomparison gap

Individual laser-scanning platforms have been evaluated for AGB estimation in many forest types; however, direct plot-level comparisons of ALS, ULS, and spaceborne waveform data under the same ecological conditions and with the same field reference are rare and absent for TDFs. Such comparisons are needed to choose platforms for operational monitoring and to understand whether denser drone data or spaceborne waveforms improve on conventional ALS. This study fills this gap for a semi-deciduous TDF using one field reference, the same plots, and an identical modelling and validation protocol for discrete-return ALS, non-repetitive-pattern ULS, and simulated full-waveform SLS.

### 1.2. Modelling approach and validation with few plots

The ordinary least squares (OLS) regression of AGB on canopy height metrics is the standard approach in laser scanning biomass studies and has a clear allometric basis: plot AGB scales with mean canopy height through a power law (Asner and Mascaro, 2014). Nonlinear methods, such as support vector machine regression (SVM), are increasingly being applied because the relationship between structure and AGB can be nonlinear in heterogeneous forests (Gleason and Im, 2012; Torre-Tojal et al., 2019; Schuh et al., 2020). In TDFs, nonlinearity is expected from successional heterogeneity, open early stands with shrubs and saplings, intermediate stands with a single deciduous canopy, and late stands with two layers and from lianas and multi-stemmed trees that add canopy volume with little stem biomass. Whether SVM can exploit these effects with the small number of plots typical of long-term TDF networks is, however, an empirical question.

With few plots, the validation design determines the reported accuracy. When candidate metrics are screened or ranked using all plots and only afterwards cross-validated, the held-out plots influence model construction, and the error is underestimated (Ambroise and McLachlan, 2002; Varma and Simon, 2006; Cawley and Talbot, 2010). The reference AGB is also uncertain because of allometry, wood density, height, and diameter errors (Chave et al., 2004, 2014; Réjou-Méchain et al., 2019), and good-practice protocols recommend quantifying this uncertainty and comparing it with the model error (Duncanson et al., 2021).

Our objectives were to (i) compute reproducible plot-level reference AGB with quantified uncertainty; (ii) compare ALS, ULS, and simulated SLS for AGB estimation; (iii) compare OLS and SVM regression under a leakage-free leave-one-plot-out protocol; and (iv) quantify how predictor-selection leakage and non-structural predictors affect apparent accuracy. This study replaces the analysis of an earlier preprint (Mattié et al., 2025), which used different reference AGB values and selected predictors before cross-validation. A companion study extends the comparison to tree-ensemble and neural-network learners and LiDAR-adapted Random Volume over Ground model.

## 2. Materials and methods

### 2.1. Study site

The study was conducted at the Santa Rosa National Park Environmental Monitoring Super Site (SRNP-EMSS), Guanacaste, Costa Rica (Fig. 1). SRNP-EMSS is a mosaic of semi-deciduous TDF in different stages of succession that was largely deforested during the 18th–20th centuries and has regenerated since the park was established in 1971 (Arroyo-Mora et al., 2005; Li et al., 2017). Stands are classified by time since abandonment into early (E), intermediate (I), and late (L) stages (Arroyo-Mora et al., 2005). Early stands consist of shrubs, small trees, and saplings up to 6–8 m tall, dominated by *Cochlospermum vitifolium*, *Gliricidia sepium*, *Rehdera trinervis,* and other heliophytic species (Hilje et al., 2015). Intermediate and late stands have a deciduous upper canopy of 10–15 m (up to 20–25 m in late stands) over a sub-canopy of lianas and shade-tolerant evergreen trees; *Rehdera trinervis* and *Guazuma ulmifolia* dominate intermediate stands, and *Calycophyllum candidissimum* and *Hymenaea courbaril* late stands (Kalacska et al., 2004) (Fig. 2).

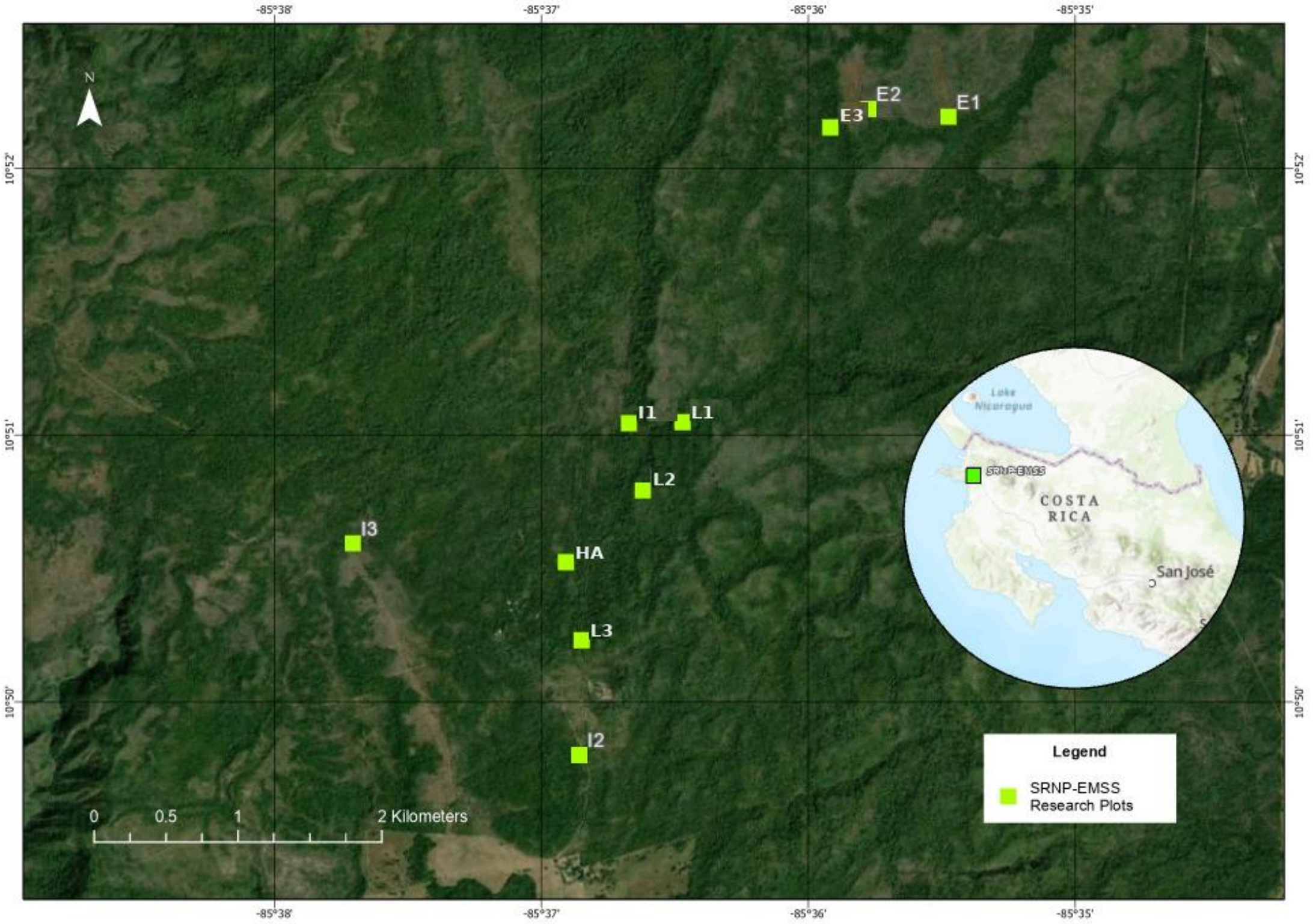

*Fig. 1. Permanent plots of the SRNP-EMSS, Guanacaste, Costa Rica. E, I and L denote early, intermediate and late successional stages; HA is the 1-ha late-stage plot. Base image: WorldView-2, Maxar, 9 January 2019. Map lines delineate study areas and do not necessarily depict accepted national boundaries.*

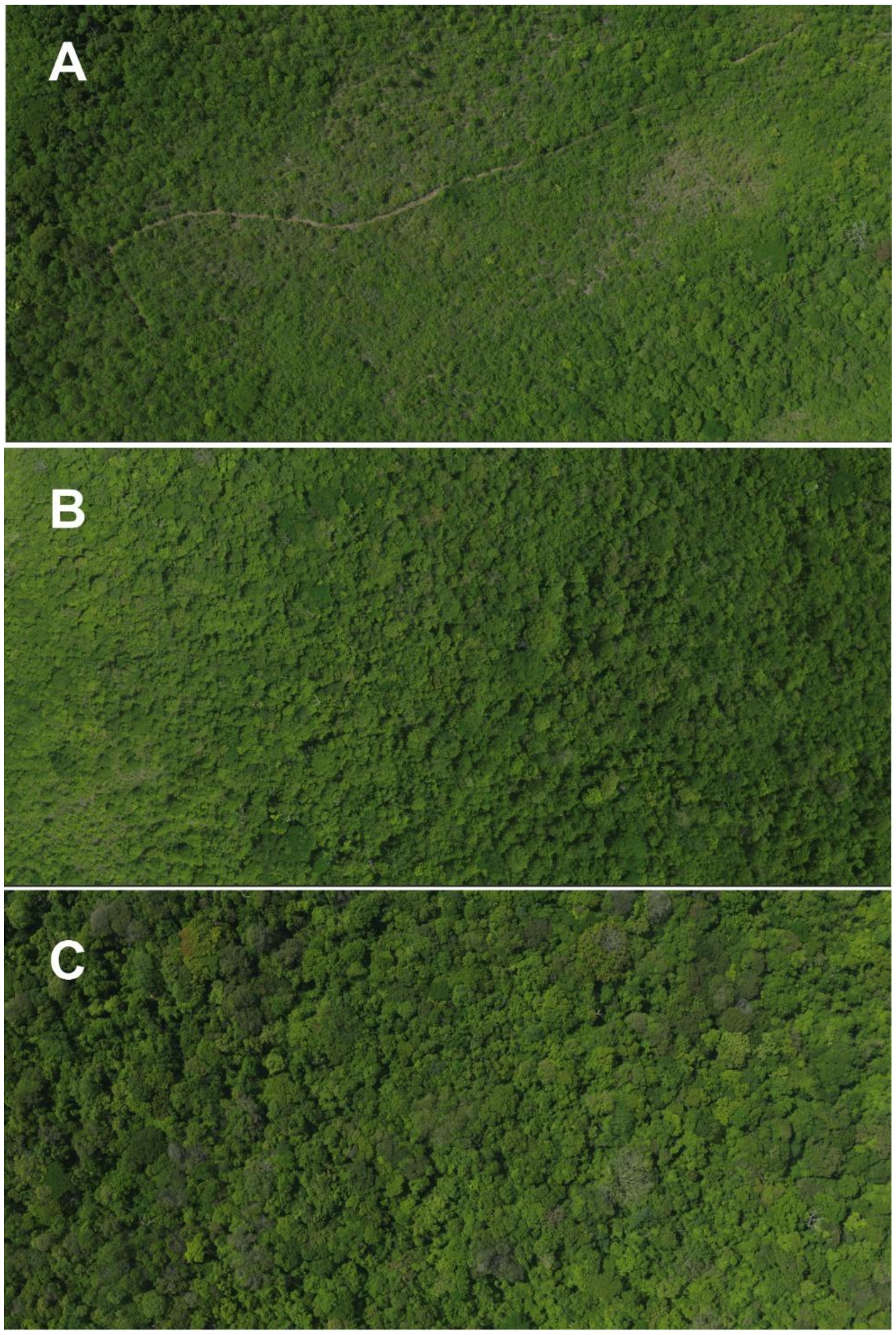

*Fig. 2. Canopy of (a) early, (b) intermediate, and (c) late successional stages at SRNP-EMSS (Hasselblad H4D-50 aerial photographs, May 2021).*

## 2.2. Field inventory and reference AGB

Nine 0.1-ha plots (20 m × 50 m; three per stage) established in 1998 and a 1-ha late-stage plot (HA) established in 2010 were re-measured annually following the Tropi-Dry protocol (Nassar et al., 2008). All stems with a diameter at breast height (DBH, 1.3 m) ≥ 5 cm were identified and measured, and tree height was recorded, and lianas were measured separately. We used the 2021 census, which brackets the laser-scanning acquisitions (March–May 2021).

Reference AGB was computed from the tree records using a single scripted procedure applied identically to every plot. Multi-stem individuals were represented by the quadratic sum of stem diameters. Diameters recorded in 2022 as circumferences were converted to DBH. A 2021 diameter was accepted when its change from the 2019 measurement was between −2 cm and +4 cm (≤ 2 cm $yr^{-1}$); otherwise, it was replaced by interpolation between 2019 and 2022 when the 2022 value was consistent, or by the 2019 value. Stems alive in 2019 and 2022 without a 2021 record were interpolated, and stems first recorded in 2022 were excluded as recruits. Missing and outlying heights were predicted from stage-specific height–diameter models ($\ln H = a + b \ln D$). Wood density (ρ) was assigned at the species level, else genus mean, else plot mean, from site data (Powers and Tiffin, 2010; Chave et al., 2009). Tree AGB (kg) was computed with the pantropical model used at SRNP-EMSS (Chave et al., 2014; Calvo-Rodriguez et al., 2021):

$$\mathrm{AGB}_{tree} = 0.0673\,(\rho\,D^2\,H)^{0.976} \qquad (1)$$

where $D$ is DBH (cm), $H$ is height (m), and ρ is wood density (g $cm^{-3}$); plot AGB (Mg $ha^{-1}$) is the sum over trees divided by plot area. Uncertainty was propagated by Monte Carlo simulation (1000 realizations) with DBH error N(0, 0.0062$D$ + 0.0904) cm (Chave et al., 2004), height and wood-density errors of 10% (SD), and the allometric residual error of Eq. 1 ($\sigma = 0.357$ on the log scale; Chave et al., 2014). The 2.5th and 97.5th percentiles give the 95% interval, and the root mean square of the plot standard deviations defines the field-uncertainty floor, the error that a perfect remote-sensing model would still show against the field data. Carbon stocks were computed with a carbon fraction of 0.47 (Calvo-Rodriguez et al., 2021):

$$C = 0.47\,\mathrm{AGB}, \qquad \mathrm{CO_2e} = \frac{44}{12}\,C = 3.67\,C \qquad (2)$$

## 2.3. Laser scanning data

**ULS.** Discrete-return data with a non-repetitive scanning pattern were collected in March 2021 with a Geosun GS-MID40 system (Livox Mid-40 laser head, 905 nm, integrated IMU/GNSS) on a DJI Matrice 300 RTK flown at 100 m above ground at 18 km $h^{-1}$. The mean return density within the plots was 776 pts $m^{-2}$ for all returns (range 400–1127) and 615 pts $m^{-2}$ for the first returns (range 366–801; Fig. 3).

**ALS.** Discrete-return data with a linear scanning pattern were collected in May 2021 using a Riegl LMS-Q680i (1550 nm, 400 kHz) integrated in an IGI LiteMapper 6800i system on a Piper PA-23 Aztec flown at 500 m above ground at 158 km $h^{-1}$. The mean return density within the plots was 59 pts $m^{-2}$ for all returns (range 34–99) and 28 pts $m^{-2}$ for the first returns (range 17–47), reflecting the overlap of flight strips.

**SLS.** All 11 GEDI shots over the plots in 2019–2021 failed the quality criteria (quality flag = 0). Therefore, we used the GEDI simulator (Hancock et al., 2019) to generate one full-waveform footprint (25 m diameter, centered on each plot) from the ALS point cloud (Fig. 4). The simulated SLS represents the spaceborne sampling geometry and waveform signal, but it is derived from the ALS acquisition and is not an independent measurement.

Trajectories were post-processed against the CORS network of Costa Rica's National Geographic Institute using AeroOffice and GrafNav, combined with IMU data. Point clouds were tiled, de-duplicated, denoised, classified into ground and non-ground points, height-normalized, and clipped to the plots using LAStools. Height metrics were computed using FUSION CloudMetrics (McGaughey, 2014), and a 0.5-m CHM was computed for every plot, from which mean canopy height ($\bar{H}$), CHM standard deviation, and 95th percentile, the fraction of the plot below 2 m, and first-return canopy cover above 1.37 m were derived. SLS metrics were computed using the simulator and rGEDI (Silva et al., 2018) (Fig. 5).

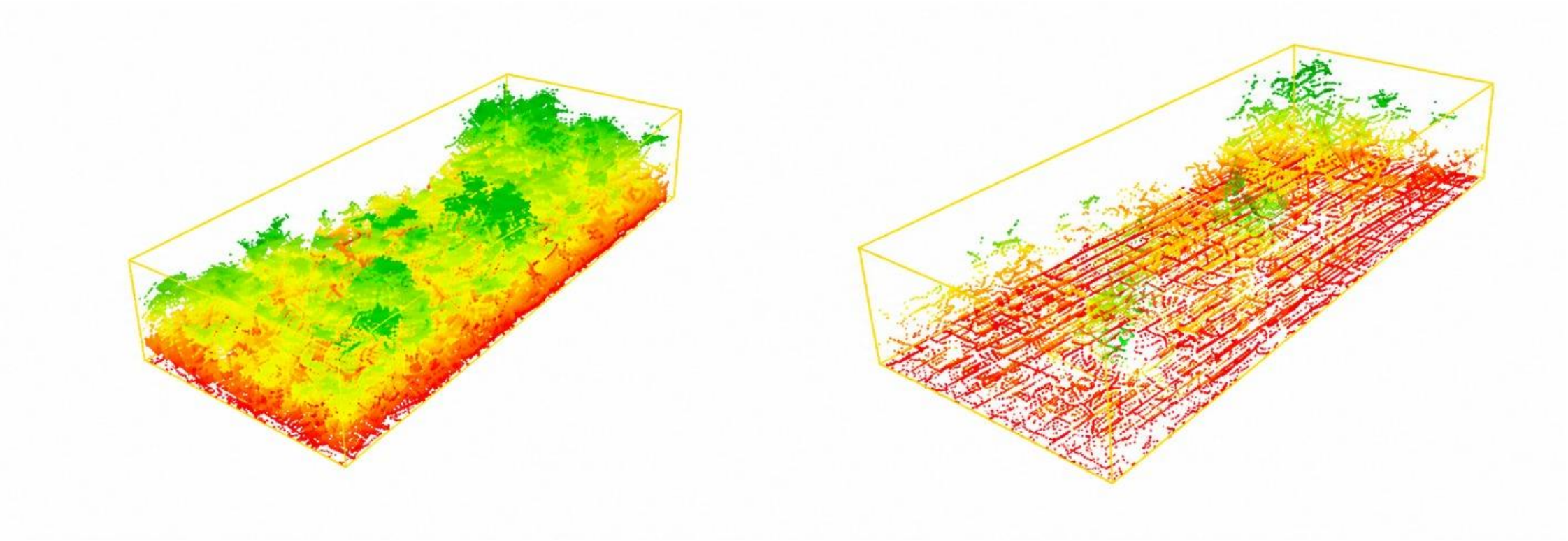

*Fig. 3. Point clouds of plot E1 from (a) ULS and (b) ALS.*

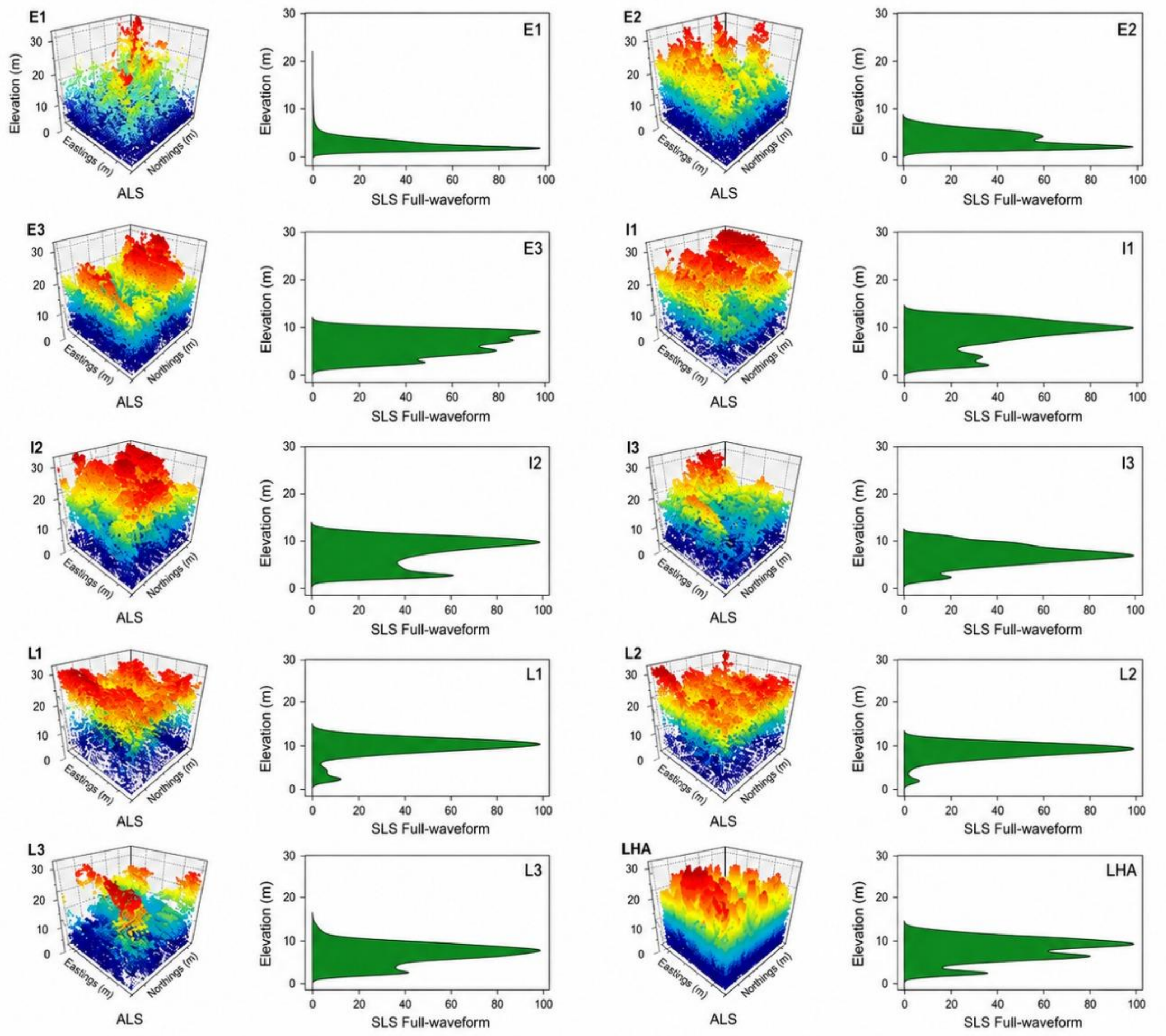

*Fig. 4. ALS point clouds and simulated GEDI (SLS) waveforms for each plot.*

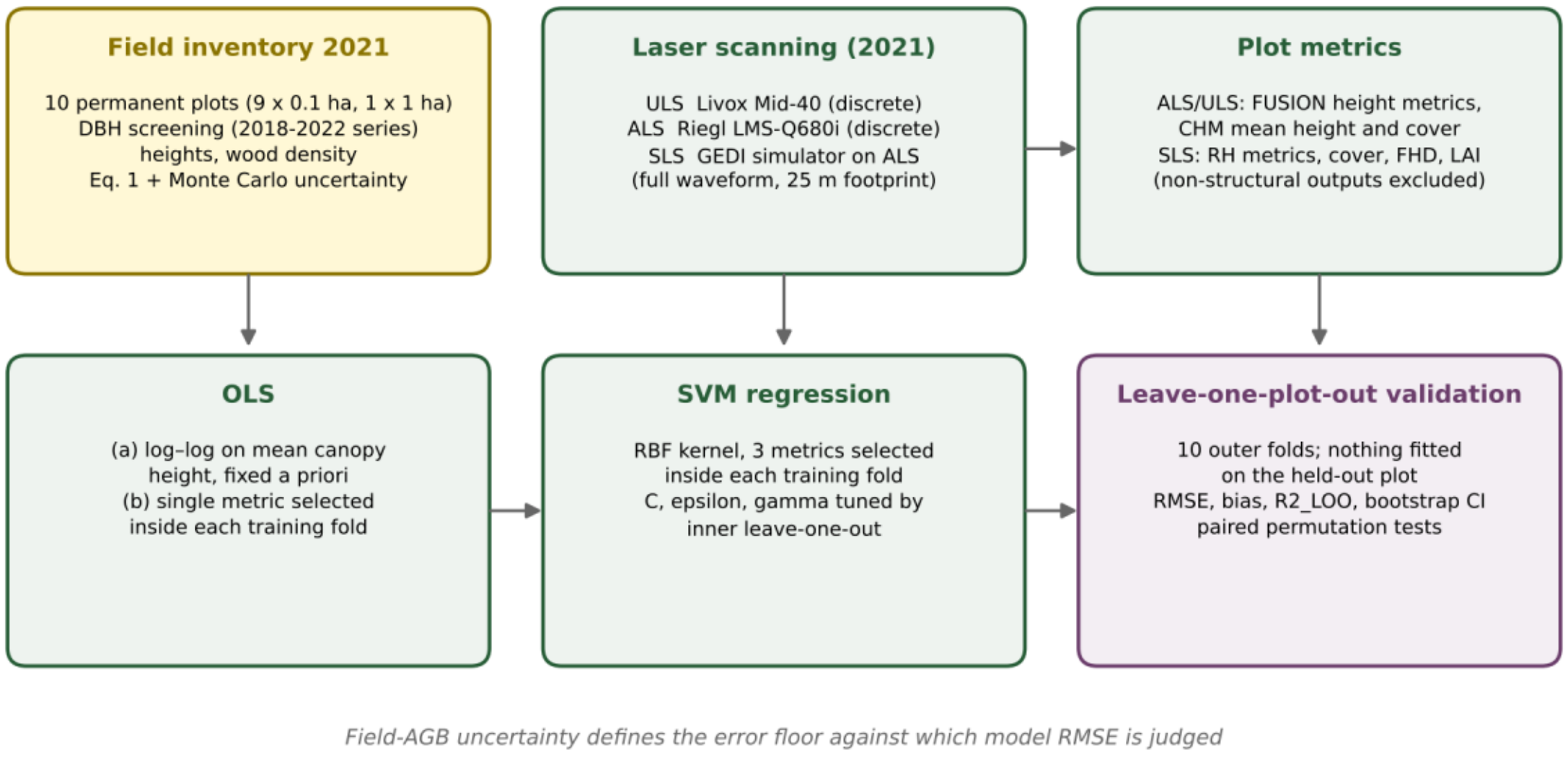

*Fig. 5. Workflow of the study. All data-driven steps were nested within the leave-one-plot-out validation.*

## 2.4. Candidate predictors

For ALS and ULS, the candidate predictors were the FUSION height statistics (mean, mode, dispersion, skewness, kurtosis, L-moments, percentiles P01–P99, canopy relief ratio, and profile area), and the CHM-derived metrics. Return counts were excluded because they scale with plot area, and the minimum normalized height was excluded because it reflects ground-classification noise. For SLS, the candidate predictors were the relative-height metrics (RH0–RH100 from Gaussian, maximum, inflection, and real ground estimates), canopy cover, leading- and trailing-edge extents, foliage height diversity, and LAI profiles. Simulator outputs that do not describe forest structure were excluded: footprint coordinates, ALS point and beam density inside the footprint, absolute ground and signal elevations, and waveform energy, which is a constant in simulations.

## 2.5. Regression models

**OLS.** Two OLS models were fitted. The first is a log–log model of mean canopy height, with the predictor fixed before analysis (mean CHM height for ALS and ULS, RH95 for SLS) and back-transformed with Baskerville's (1972) correction:

$$\ln\text{AGB} = \beta_0 + \beta_1 \ln\bar{H} + \varepsilon, \qquad \widehat{\text{AGB}} = \exp\left(\hat{\beta}_0 + \hat{\beta}_1 \ln\bar{H} + \hat{\sigma}^2/2\right) \qquad (3)$$

The second was a linear model of the single metric most correlated with AGB (absolute Spearman correlation) in each training fold.

**SVM regression.** SVM regression estimates a function that deviates from the observed AGB by at most ε while remaining as flat as possible, with a radial basis kernel representing nonlinear relationships without specifying a functional form (Smola and Schölkopf, 2004):

$$\hat{f}(\mathbf{x}) = \sum_{i=1}^{n} (\alpha_i - \alpha_i^*) \exp(-\gamma \parallel \mathbf{x}_i - \mathbf{x} \parallel^2) + b \qquad (4)$$

where $\alpha_i$ and $\alpha_i^*$ are the dual coefficients (non-zero for support vectors), γ is the kernel width, and *b* is the bias. In each training fold, the three metrics most correlated with AGB were retained after discarding any metric correlated at $|r| > 0.95$ with one already chosen; predictors were standardized with training-fold statistics, and $C \in \{10, 100, 1000\}$, $\varepsilon \in \{1, 5\}$ Mg ha$^{-1}$ and $\gamma \in \{0.05, 0.2, 1\}$ were tuned by an inner leave-one-out loop.

### 2.6. Validation and comparison

All models were evaluated using leave-one-plot-out cross-validation (LOO, 10 folds), the most data-efficient scheme for ten plots. Predictor selection, standardization, and hyper-parameter tuning used only the nine training plots of each fold so that the held-out plot never informed its own prediction (Varma and Simon, 2006; Roberts et al., 2017). From the ten held-out predictions we computed

$$\mathrm{RMSE} = \sqrt{1/n\sum_i (\hat{y}_i - y_i)^2}, \quad \mathrm{MAE} = 1/n\sum_i |\hat{y}_i - y_i|, \quad \mathrm{Bias} = 1/n\sum_i (\hat{y}_i - y_i), \quad R^2_{LOO} = 1 - \frac{\sum_i (\hat{y}_i - y_i)^2}{\sum_i (y_i - \bar{y})^2} \tag{5}$$

and the relative RMSE (rRMSE, % of mean field AGB). $R^2_{LOO}$ applies to both linear and nonlinear models. Ninety-five percent intervals of RMSE were obtained by bootstrapping the plot errors (2000 resamples). Models and platforms were compared with two-sided sign-flip permutation tests on paired squared errors (20 000 permutations). To quantify selection bias, SVM was also run with three predictors selected once on all ten plots and with the SLS predictor set of the earlier preprint. Analyses were performed in Python (scikit-learn, NumPy, SciPy); data and code are available from the corresponding author upon reasonable request (see Data and code availability).

## 3. Results

### 3.1. Reference AGB

Reference AGB ranged from 21.8 Mg ha$^{-1}$ (E1) to 220.3 Mg ha$^{-1}$ (L1), with a mean of 111.0 Mg ha$^{-1}$, corresponding to 10.2–103.5 Mg C ha$^{-1}$ (Table 1, Fig. 6). The 95% Monte Carlo intervals were 3.5–14.5% of plot AGB, smallest in the 1-ha plot. Trees with corrected 2021 diameters held 0–35% of plot AGB. The field-uncertainty floor was 11.2 Mg ha$^{-1}$ (10% of mean AGB). AGB was low in the early plots, highest in I1, L1 and L2, and intermediate in I2, I3, L3 and the 1-ha plot, showing that successional stage alone does not determine biomass.

Table 1. Permanent plots, reference AGB, and carbon stock (census 2021).

| Plot | Stage | Area (ha) | Trees (DBH ≥ 5 cm) | Basal area ($m^2$ $ha^{-1}$) | Mean canopy height, ALS (m) | AGB (Mg $ha^{-1}$) [95% interval] | Carbon (Mg C $ha^{-1}$) | Uncertainty (%) |
|---|---|---|---|---|---|---|---|---|
| E1 | Early | 0.1 | 113 | 8.1 | 3.1 | 21.8 [19.4–25.1] | 10.2 | 6.6 |
| E2 | Early | 0.1 | 81 | 9.1 | 5.3 | 23.9 [20.4–28.4] | 11.2 | 8.4 |
| E3 | Early | 0.1 | 97 | 14.3 | 9.0 | 61.4 [51.6–75.0] | 28.8 | 9.8 |
| I1 | Intermediate | 0.1 | 105 | 37.2 | 15.0 | 204.6 [174.8–239.2] | 96.1 | 8.4 |
| I2 | Intermediate | 0.1 | 58 | 17.7 | 12.4 | 77.4 [60.0–103.0] | 36.4 | 14.2 |
| I3 | Intermediate | 0.1 | 66 | 17.1 | 8.9 | 75.7 [65.1–88.9] | 35.6 | 8.4 |
| L1 | Late | 0.1 | 132 | 36.4 | 15.4 | 220.3 [188.8–262.5] | 103.5 | 8.8 |
| L2 | Late | 0.1 | 134 | 35.5 | 12.9 | 174.8 [154.2–203.9] | 82.1 | 7.1 |
| L3 | Late | 0.1 | 84 | 24.3 | 9.7 | 114.0 [88.3–151.4] | 53.6 | 14.5 |
| HA | Late | 1.0 | 726 | 22.5 | 13.2 | 136.1 [127.5–146.1] | 63.9 | 3.5 |

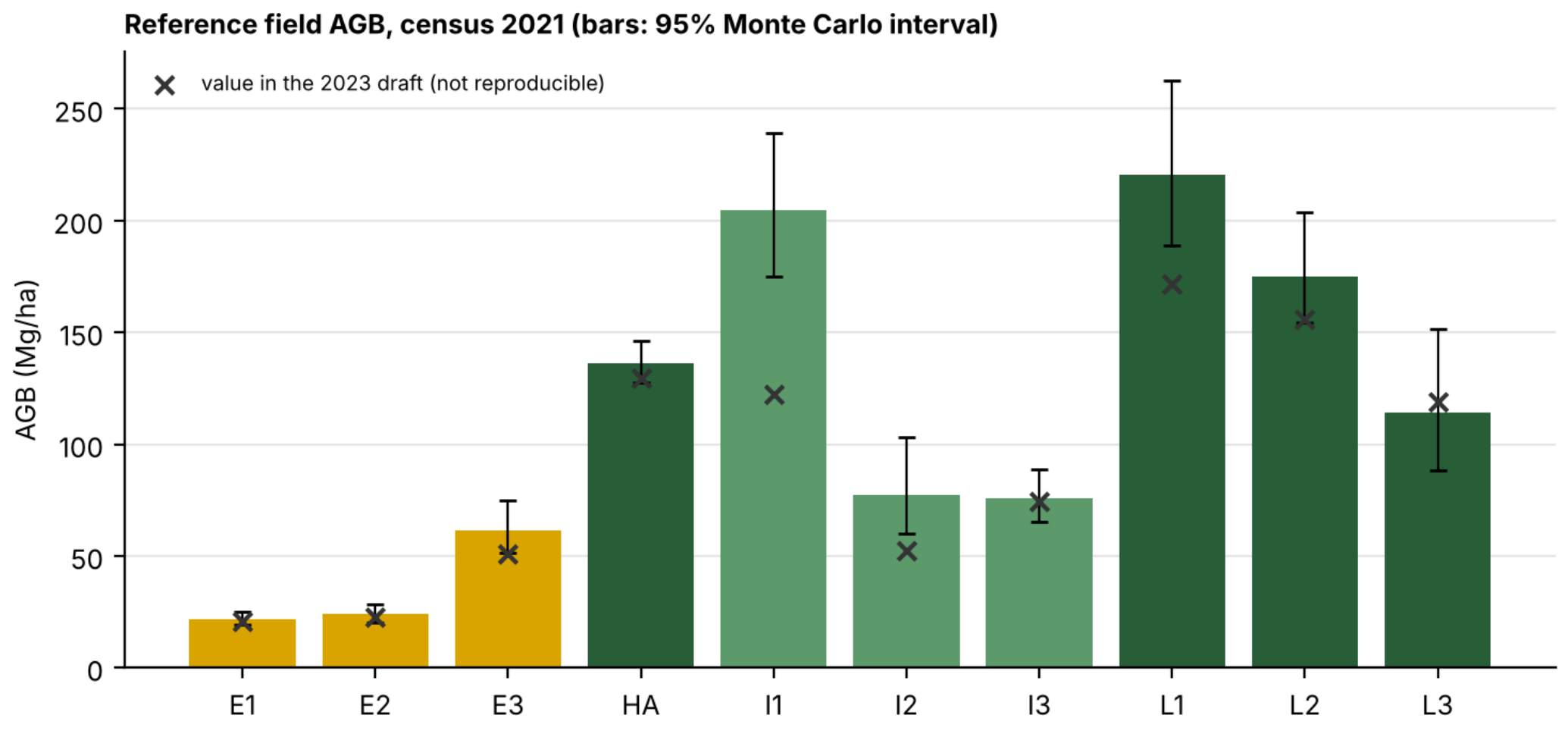

*Fig. 6. Reference field AGB (census 2021) with 95% Monte Carlo intervals. The crosses show the values used in the earlier preprint, which could not be reproduced from the tree records.*

## 3.2. Selected predictors

For ALS, the single-metric OLS model selected the FUSION mean height in every fold, and SVM combined mean height with distribution-shape metrics (Table 2). For ULS, mid-canopy percentiles (P40) were selected most often. For SLS, relative heights of the lower and middle canopy (RH30–RH55 from the inflection-point ground) were selected rather than top-of-canopy heights. Therefore, height metrics carried the AGB signal for all platforms.

Table 2. Metrics selected inside the training folds (number of folds out of 10 in parentheses).

| Platform | OLS: metric selected (folds out of 10) | SVM: most frequent metrics (folds out of 10) |
|---|---|---|
| ALS | Elev mean (10) | Elev mean (10), Elev L4 (10), Elev L kurtosis (2), Canopy relief ratio (2) |
| ULS | Elev P40 (6), Elev mean (2), Elev mode (1) | Elev L kurtosis (6), Elev P40 (6), Elev P05 (4), Elev L4 (3) |
| SLS | rhInfl 30 (3), rhInfl 55 (3), rhMax 35 (1) | rhInfl 30 (5), rhInfl 55 (5), rhMax 35 (3), rhMax 50 (3) |

## 3.3. OLS versus SVM

Table 3 and Figs. 7–8 summarize the leave-one-plot-out results. With ALS, the a priori log–log OLS model explained 77% of the out-of-sample variance with negligible bias (-1.8 Mg $ha^{-1}$), the single-metric OLS performed similarly (32.6 Mg $ha^{-1}$ (29%)), and SVM had a slightly larger error (35.8 Mg $ha^{-1}$ (32%); $p = 0.59$ against the a priori OLS). With ULS, the single-metric OLS was the best model (36.3 Mg $ha^{-1}$ (33%) and SVM was significantly worse (54.6 Mg $ha^{-1}$ (49%); $p = 0.01$). With SLS, the single-metric OLS reached 41.1 Mg $ha^{-1}$ (37%) and SVM 51.9 Mg $ha^{-1}$ (47%), while the a priori RH95 model performed poorly ($R^2_{LOO} = 0.18$). SVM therefore did not outperform OLS for any platform. The single-metric linear OLS predicted a negative AGB for plot E1 with ALS, which the log–log form avoided.

For all models and platforms, plot I2 was overestimated, and the highest-biomass plots I1 and L1 were underestimated (Fig. 8). All errors were at least 2.5 times larger than the field-uncertainty floor (Fig. 7).

Table 3. Leave-one-plot-out accuracy (n = 10 plots; mean field AGB = 111.0 Mg $ha^{-1}$). *p*: Two-sided permutation test of paired squared errors of SVM against the a priori OLS / the single-metric OLS.

| Platform | Model | RMSE (Mg $ha^{-1}$) [95% CI] | rRMSE (%) | MAE (Mg $ha^{-1}$) | Bias (Mg $ha^{-1}$) | $R^2_{LOO}$ | *p*, SVM vs OLS |
|---|---|---|---|---|---|---|---|
| ALS | OLS log–log, height fixed a priori | 32.3 [20.3–43.3] | 29.1 | 27.7 | -1.8 | 0.77 | – |
| ALS | OLS, single metric selected in fold | 32.6 [19.3–44.6] | 29.4 | 27.1 | -2.7 | 0.77 | – |
| ALS | SVM regression (RBF kernel) | 35.8 [28.4–42.6] | 32.3 | 33.8 | -6.3 | 0.72 | 0.59 / 0.66 |
| ULS | OLS log–log, height fixed a priori | 44.3 [18.1–65.1] | 39.9 | 30.4 | +0.2 | 0.57 | – |
| ULS | OLS, single metric selected in fold | 36.3 [18.7–53.2] | 32.7 | 28.1 | +2.4 | 0.71 | – |
| ULS | SVM regression (RBF kernel) | 54.6 [34.7–73.4] | 49.2 | 45.7 | -5.6 | 0.35 | 0.03 / 0.01 |
| SLS | OLS log–log, height fixed a priori | 61.2 [35.3–80.5] | 55.1 | 46.7 | +5.5 | 0.18 | – |
| SLS | OLS, single metric selected in fold | 41.1 [25.8–54.3] | 37.0 | 34.3 | +1.2 | 0.63 | – |
| SLS | SVM regression (RBF kernel) | 51.9 [23.7–74.3] | 46.8 | 37.4 | -5.0 | 0.41 | 0.34 / 0.29 |

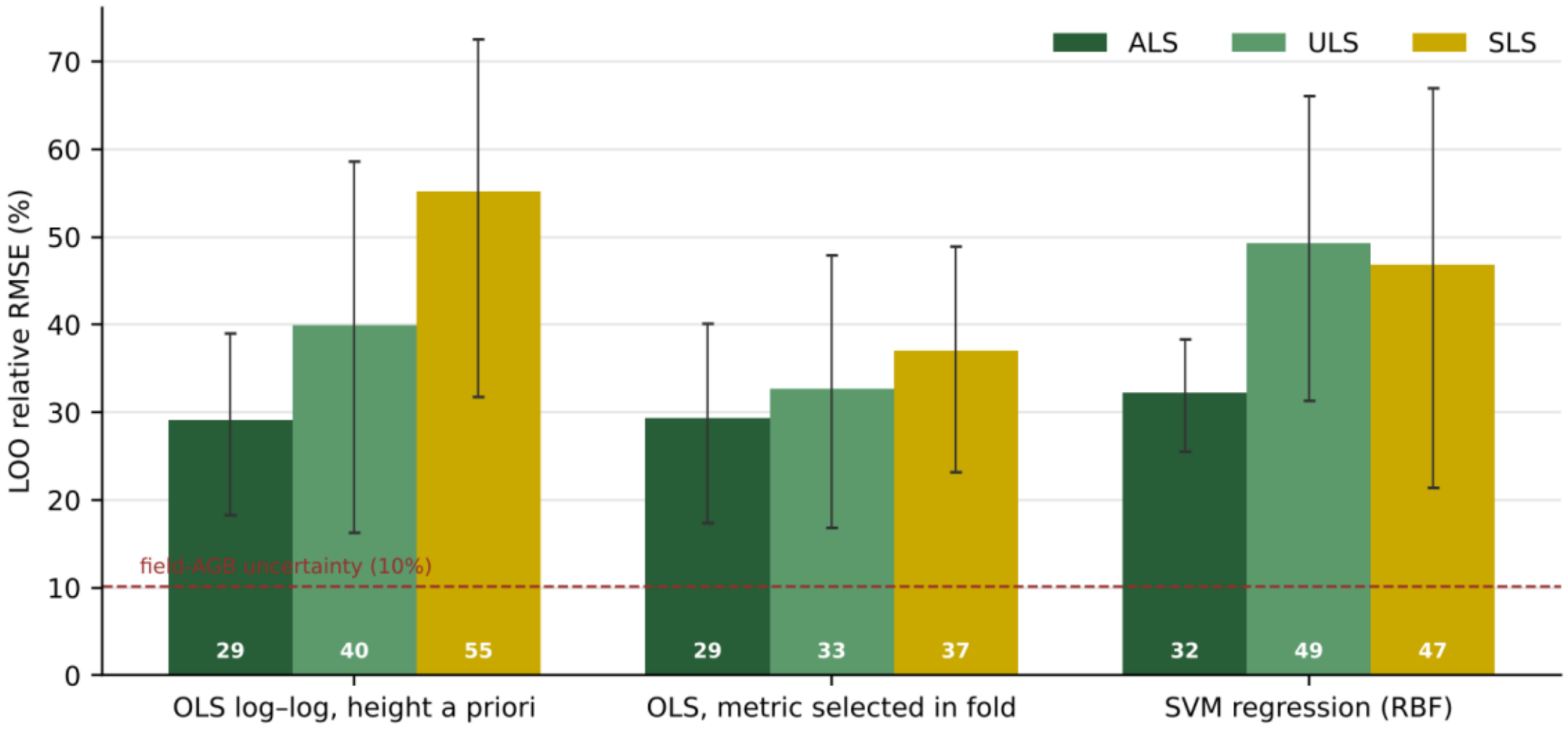

*Fig. 7. Leave-one-plot-out relative RMSE of OLS and SVM models for ALS, ULS, and SLS. Bars show 95% bootstrap intervals; the dashed line is the field-AGB uncertainty floor.*

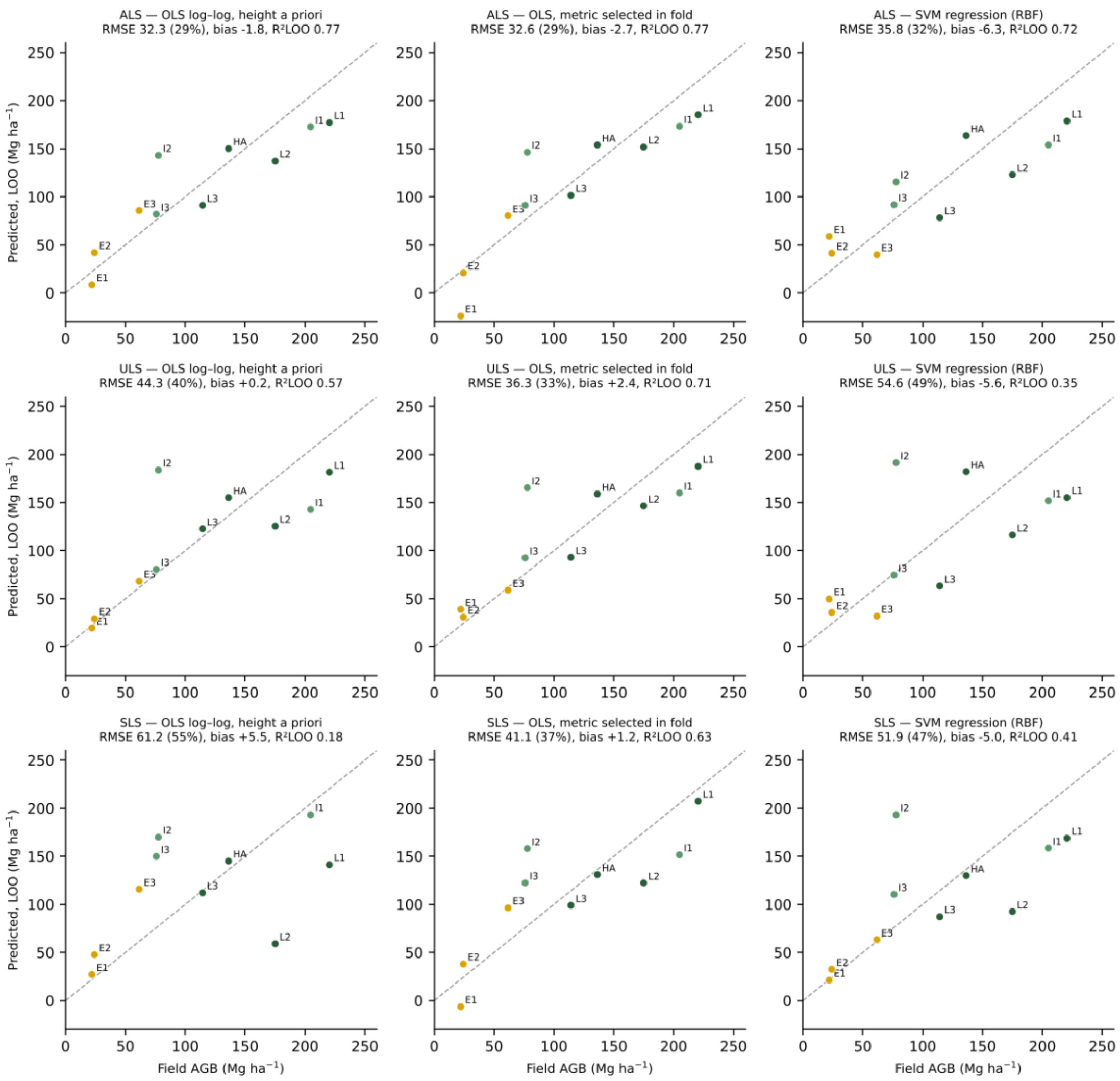

*Fig. 8. Observed field AGB versus leave-one-plot-out predictions (E early, I intermediate, and L/HA late plots).*

## 3.4. Comparison of platforms

ALS had the lowest error for the a priori OLS and SVM models, and ALS and ULS performed similarly to the single-metric OLS. However, the differences between the platforms were not statistically significant for most models (Table 4); only SVM with ULS was marginally worse than SVM with ALS.

Table 4. Difference in RMSE of ULS and SLS relative to ALS for the same model and permutation *p*-values.

| **Model** | **ULS vs ALS: ΔRMSE (Mg ha$^{-1}$), *p*** | **SLS vs ALS: ΔRMSE (Mg ha$^{-1}$), *p*** |
|---|---|---|
| OLS log–log, height fixed a priori | +12.0, 0.32 | +28.9, 0.06 |
| OLS, single metric selected in fold | +3.7, 0.56 | +8.5, 0.16 |
| SVM regression (RBF kernel) | +18.8, 0.05 | +16.1, 0.40 |

### 3.5. Effect of predictor-selection leakage

Selecting the SVM predictors once on all ten plots markedly changed the apparent accuracy (Table 5, Fig. 9). For ULS and SLS, leaky selection lowered the rRMSE from 49% to 27% and from 47% to 32%, respectively; for ALS, where mean height dominated the ranking in every fold, both protocols gave similar errors. The SLS predictor set of the earlier preprint (footprint latitude, ALS point density, ground elevations, Blair's sensitivity, and LAI) gave an rRMSE of 81% against the reference AGB.

Table 5. SVM accuracy with predictor selection outside and inside the validation loop

| Platform | Predictor selection | Predictors | RMSE (Mg ha$^{-1}$) | rRMSE (%) | $R^2_{LOO}$ |
|---|---|---|---|---|---|
| ALS | on all 10 plots, before CV | Elev mean, Elev L4, Elev P95 | 40.0 | 36.1 | 0.65 |
| ALS | inside each training fold | fold-specific | 35.8 | 32.3 | 0.72 |
| ULS | on all 10 plots, before CV | Elev P40, Elev L kurtosis, Elev P05 | 29.6 | 26.7 | 0.81 |
| ULS | inside each training fold | fold-specific | 54.6 | 49.2 | 0.35 |
| SLS | on all 10 plots, before CV | rhInfl 30, rhMax 60, rhInfl 65 | 35.5 | 32.0 | 0.72 |
| SLS | inside each training fold | fold-specific | 51.9 | 46.8 | 0.41 |
| SLS | SLS predictors of the 2025 preprint | lat, pointDense, maxGround, inflGround, blairSense, gLAI0t10 | 89.9 | 81.0 | -0.77 |

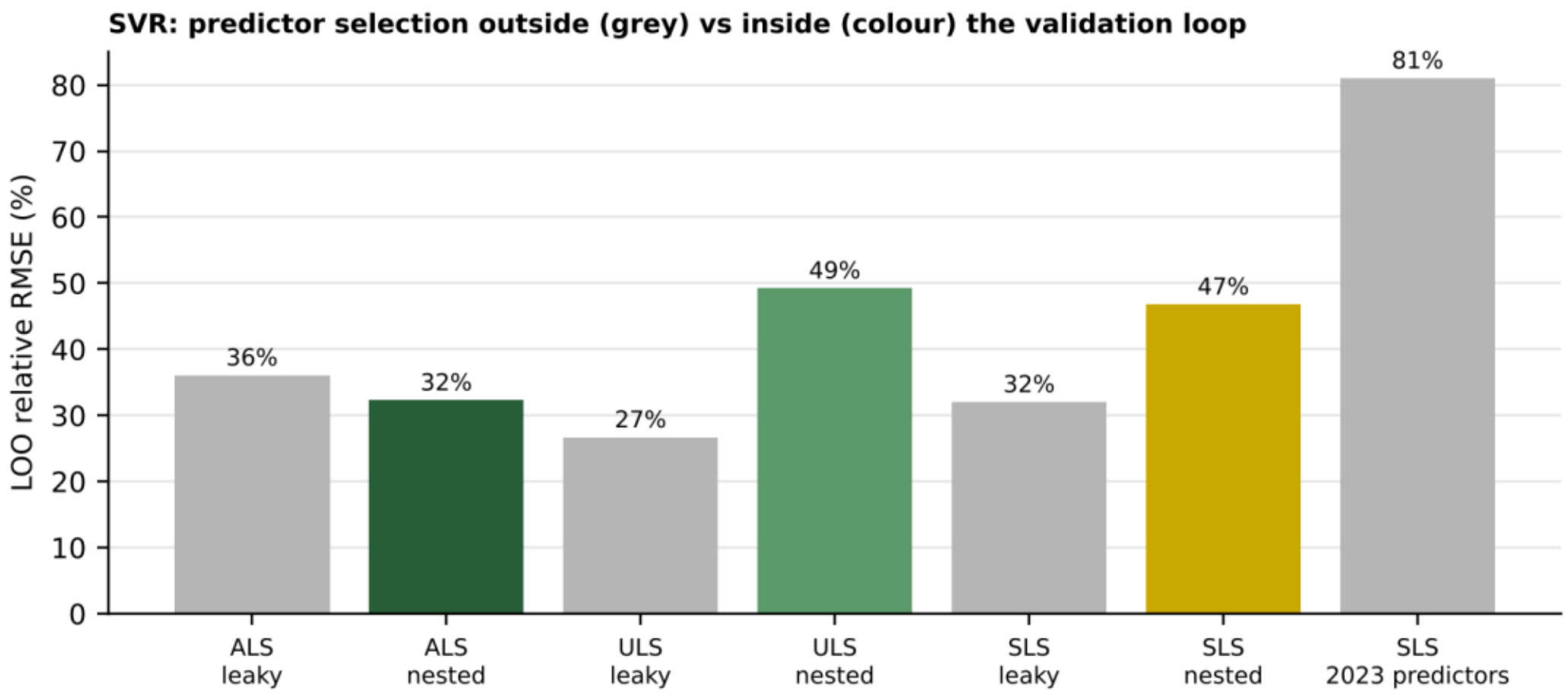

*Fig. 9. Apparent SVM accuracy when predictors are selected on all plots (grey) or inside each training fold (color), and for the SLS predictor set of the earlier preprint.*

### 3.6. Sensitivity to the field reference

Because several plot AGB values changed when the diameter series were screened, all models were also evaluated against the plot values of the earlier preprint (Table 6). The conclusions were unchanged: ALS gave the lowest errors for the a priori OLS model, SVM was not significantly more accurate than OLS with ALS ($p$ = 0.50), and remained significantly worse with ULS ($p$ = 0.04). With the earlier values, the single-metric OLS with ALS selected the fourth L-moment of the height distribution in nine of ten folds, a statistically convenient but ecologically uninterpretable metric, which illustrates why a predictor fixed a priori is preferable with few plots.

Table 6. Leave-one-plot-out accuracy with the reference AGB and earlier plot values

| Platform | Model | rRMSE, reference (%) | rRMSE, earlier values (%) | $R^2_{LOO}$, reference | $R^2_{LOO}$, earlier values |
|---|---|---|---|---|---|
| ALS | OLS log–log, height fixed a priori | 29.1 | 36.2 | 0.77 | 0.59 |
| ALS | OLS, single metric selected in fold | 29.4 | 23.7 | 0.77 | 0.82 |
| ALS | SVM regression (RBF kernel) | 32.3 | 28.5 | 0.72 | 0.75 |
| ULS | OLS log–log, height fixed a priori | 39.9 | 40.6 | 0.57 | 0.48 |
| ULS | OLS, single metric selected in fold | 32.7 | 40.0 | 0.71 | 0.50 |
| ULS | SVM regression (RBF kernel) | 49.2 | 49.2 | 0.35 | 0.24 |
| SLS | OLS log–log, height fixed a priori | 55.1 | 59.0 | 0.18 | -0.09 |
| SLS | OLS, single metric selected in fold | 37.0 | 45.3 | 0.63 | 0.36 |
| SLS | SVM regression (RBF kernel) | 46.8 | 39.4 | 0.41 | 0.51 |

# 4. Discussion

## 4.1. SVM did not improve on OLS

The ecological arguments for nonlinear modelling in TDFs, successional heterogeneity, multi-layered canopies, and lianas remain valid, but with ten plots SVM could not exploit them. The kernel model has more parameters to tune than there are plots to constrain them; therefore, its predictions vary from fold to fold, whereas the log–log OLS model has two parameters and an allometric basis (Asner and Mascaro, 2014). Most of the nonlinearity between height and AGB is captured by the logarithmic transformation, which also keeps predictions positive. Our results agree with the recommendations to restrict laser-scanning biomass models to one to four well-chosen predictors when calibration data are limited (Stahl et al., 2011; Duncanson et al., 2022) and show that the earlier conclusion that SVM substantially outperformed OLS (Mattié et al., 2025) arose from in-sample error measures, an OLS model with up to seven predictors, and predictor selection outside cross-validation.

## 4.2. ALS, ULS and simulated spaceborne data

ALS recorded the lowest errors, although its density (59 pts $m^{-2}$) was thirteen times lower than that of ULS for all returns. Plot-level AGB depends mainly on canopy height and cover, which moderate densities already resolve (Silva et al., 2017). ULS mean canopy heights exceeded ALS heights by up to 2.7 m, probably reflecting the different sensitivity of the 905-nm Livox sensor and its non-repetitive pattern to small branches and foliage, the two-month difference between acquisitions at the onset of the rainy season, and ground classification under dense understorey. These differences add noise without adding information about biomass.

The simulated spaceborne waveforms were the least accurate for structural and geometric reasons. A single 25-m footprint covers approximately 49% of a 0.1-ha plot and 5% of a 1-ha plot, and the waveform is derived from the same ALS points; therefore, it cannot add independent information. The earlier finding that SLS was the most accurate platform (Mattié et al., 2025) was produced by predictors that do not describe forest structure, footprint coordinates, which separate the spatially clustered successional stages, ALS point density, and terrain elevation, selected using all plots. The selection of lower- and mid-canopy RH metrics in the present analysis suggests that full-waveform data capture the vertical distribution of vegetation, but exploiting this information with real GEDI data in TDFs will require several high-quality footprints per calibration unit and plots larger than the footprint (Duncanson et al., 2021, 2022).

### 4.3. Sample size and plot size

Ten plots are a small sample, and the 0.1-ha plots are smaller than the plot sizes recommended for calibrating and validating biomass products (Duncanson et al., 2021). With ten plots, bootstrap intervals of RMSE span 20–40 Mg ha$^{-1}$, and only large differences between models or platforms can be detected. Small plots also increase the influence of edge effects and of single large trees on plot AGB (Mascaro et al., 2011; Réjou-Méchain et al., 2019). We addressed these limitations with a leakage-free leave-one-plot-out protocol, models with one to three predictors, bootstrap intervals, and permutation tests; however, the results should be regarded as a case study that motivates larger networks rather than as transferable models.

### 4.4. Field AGB uncertainty

Field AGB uncertainty (3.5–14.5% per plot, 10% overall) was small compared with model error; therefore, the remote-sensing models, rather than field data, limit accuracy at present. Nevertheless, reference values changed substantially when the diameter series were screened, particularly in plots I1, I2, and L1, where a few large stems dominated AGB. Therefore, publishing the tree-level calculation used to derive reference AGB is as important as reporting model accuracy; the sensitivity analysis (Section 3.6) shows that our comparative conclusions do not depend on the set of plot values used, although absolute errors do. Plot I2, overestimated by every model, combines a tall, open canopy with relatively few large trees; whether this reflects a liana-dominated canopy, measurement problems, or genuine structural variability should be resolved in the field.

### 4.5. Implications and future research

For laser-scanning AGB estimation in TDFs with small plot networks, ALS mean canopy height in a log–log OLS model provides the most robust option, and every data-driven step, including predictor screening, must be nested within cross-validation. Future work should (i) enlarge the calibration set with larger plots at SRNP-EMSS and other dry forests; (ii) use high-quality GEDI footprints co-located with field plots; (iii) compare additional learners and physically based models, as done in our companion study; and (iv) once larger training sets allow complex models, use explainable-AI methods for 3D point clouds, such as counterfactual explanations (Holzinger et al., 2025), to show which changes in canopy height distribution or waveform shape drive the predicted biomass and to increase confidence in model outputs for carbon reporting.

## 5. Conclusions

This study presents a plot-level intercomparison of airborne, unmanned, and simulated spaceborne laser scanning for AGB estimation in a tropical dry forest, with a reproducible field reference and leakage-free validation. (1) ALS gave the lowest errors (about 29% rRMSE, $R^2_{LOO}$ ≈ 0.77), although differences between platforms were not statistically significant with ten plots. (2) SVM regression did not improve on OLS for any platform and was significantly worse with ULS; a log–log OLS model of mean canopy height is the most robust option for small plot networks. (3) Thirteen-fold denser ULS data and simulated GEDI waveforms did not improve accuracy. (4) Selecting predictors on all plots and using non-structural simulator outputs inflated apparent accuracy, which explains the higher accuracies reported in an earlier version of this work. (5) Model errors were at least 2.5 times larger than the field-AGB uncertainty, and the comparative conclusions held with an earlier set of field values. Larger and more numerous plots, and co-located high-quality spaceborne footprints, are needed before nonlinear models can be shown to improve biomass estimates in tropical dry forests.

## CRediT authorship contribution statement

**Nelson Mattié:** Conceptualization, Methodology, Software, Formal analysis, Investigation, Data curation, Visualization, Writing – original draft. **Arturo Sanchez-Azofeifa:** Conceptualization, Methodology, Investigation, Resources, Supervision, Writing – review & editing. **Pablo Crespo-Peremarch:** Conceptualization, Supervision, Writing – review & editing. **Juan-Ygnacio López-Hernández:** Methodology, Software, Formal analysis, Visualization, Writing – review & editing.

## Declaration of competing interest

The authors declare that they have no known competing financial interests or personal relationships that could have influenced the work reported in this paper.

## Data and code availability

The tree-level census data of the SRNP-EMSS permanent plots and the derived plot-level data are held by the University of Alberta and cannot be shared publicly without authorization from the data owners. Data and modelling code are available from the corresponding author on reasonable request, subject to that authorization.

## Funding

This research did not receive any specific grants from funding agencies in the public, commercial, or not-for-profit sectors.

## Acknowledgements

The authors acknowledge the support of the Centre for Earth Observation Sciences, Department of Earth and Atmospheric Sciences, University of Alberta; the Earth Observation Center Hemera, Universidad Mayor, Chile; and Natural Resources Canada. We thank NASA for the GEDI data and simulator and Osvaldo Valeria (Université du Québec en Abitibi-Témiscamingue) for comments on an early version of the manuscript.

## Declaration of generative AI and AI-assisted technologies in the manuscript preparation process

During the preparation of this work, the authors used Claude (Anthropic) to assist with editing the text and with the analysis code. After using this tool, the authors reviewed and edited the content as needed and take full responsibility for the content of the published article.